\documentclass[aps,prl,twocolumn,amsmath,amssymb,showpacs,superscriptaddress,notitlepage,longbibliography]{revtex4-1}
\usepackage[colorlinks=true,linkcolor=blue,anchorcolor=red,citecolor=blue, urlcolor=blue]{hyperref}
\usepackage{bm}
\usepackage{graphicx}
\usepackage{color}
\usepackage{subfigure}
\usepackage{amsmath, bm}

\begin{document}

\title{Quantum theory of the nonlinear Hall effect}

\author{Z. Z. Du}
\affiliation{Shenzhen Institute for Quantum Science and Engineering and Department of Physics, Southern University of Science and Technology (SUSTech), Shenzhen 518055, China}
\affiliation{Shenzhen Key Laboratory of Quantum Science and Engineering, Shenzhen 518055, China}

\author{C. M. Wang}
\affiliation{Department of Physics, Shanghai Normal University, Shanghai 200234, China}
\affiliation{Shenzhen Institute for Quantum Science and Engineering and Department of Physics, Southern University of Science and Technology (SUSTech), Shenzhen 518055, China}
\affiliation{Shenzhen Key Laboratory of Quantum Science and Engineering, Shenzhen 518055, China}

\author{Hai-Peng Sun}
\affiliation{Shenzhen Institute for Quantum Science and Engineering and Department of Physics, Southern University of Science and Technology (SUSTech), Shenzhen 518055, China}
\affiliation{Shenzhen Key Laboratory of Quantum Science and Engineering, Shenzhen 518055, China}

\author{Hai-Zhou Lu}
\email{Corresponding author: luhz@sustech.edu.cn}
\affiliation{Shenzhen Institute for Quantum Science and Engineering and Department of Physics, Southern University of Science and Technology (SUSTech), Shenzhen 518055, China}
\affiliation{Shenzhen Key Laboratory of Quantum Science and Engineering, Shenzhen 518055, China}

\author{X. C. Xie}
\affiliation{International Center for Quantum Materials, School of Physics, Peking University, Beijing 100871, China}
\affiliation{CAS Center for Excellence in Topological Quantum Computation, University of Chinese Academy of Sciences, Beijing 100190, China}
\affiliation{Beijing Academy of Quantum Information Sciences, West Building 3, No.10, Xibeiwang East Road, Haidian District, Beijing 100193, China}

\date{\today }

\begin{abstract}
The nonlinear Hall effect is an unconventional response, in which a voltage can be driven by two perpendicular currents in the Hall-bar measurement.
Unprecedented in the family of the Hall effects, it can survive time-reversal symmetry but is sensitive to the breaking of discrete and crystal symmetries.
It is a quantum transport phenomenon that has deep connection with the Berry curvature.
However, a full quantum description is still absent. Here we construct a quantum theory of the nonlinear Hall effect by using the diagrammatic technique.
Quite different from nonlinear optics, nearly all the diagrams account for the disorder effects, which play decisive role in the electronic transport.
After including the disorder contributions in terms of the Feynman diagrams, the total nonlinear Hall conductivity is enhanced but its sign remains unchanged for the 2D tilted Dirac model, compared to the one with only the Berry curvature contribution.
We discuss the symmetry of the nonlinear conductivity tensor and predict a pure disorder-induced nonlinear Hall effect for point groups $C_{3}$, $C_{3h}$, $C_{3v}$, $D_{3h}$, $D_{3}$ in 2D, and $T$, $T_{d}$, $C_{3h}$, $D_{3h}$ in 3D. This work will be helpful for explorations of the topological physics beyond the linear regime.
\end{abstract}

\maketitle

The recently discovered nonlinear Hall effect \cite{Sodemann15prl,Low15prb,Facio18prl,You18prb,ZhangY18-2dm,Du18prl,Ma19nat,Kang19nm,Du19nc,Xiao19prb,Nandy19prb,Matsyshyn19prl,WangH19npjqm,Shvetsov19jetp,Son2019PRL,ZhouT20prap,Habib20prr,Shao20prl,Singh20PRL,Tu202D,LiZ20arXiv,Bhalla20PRL}
is a new member of the Hall family \cite{Klitzing80prl,Prange12book,Nagaosa10rmp,Yasuda16np}.
It is characterized by a nonlinear transverse voltage (or current) in response to two $ac$ currents (or electric fields).
The nonlinear Hall effect does not require breaking time-reversal symmetry but inversion symmetry.
More importantly, this effect is an unconventional response sensitive to the breaking of discrete
and crystal symmetries, and thus can be used to probe phase transitions induced by spontaneous symmetry breaking, such as ferroelectric \cite{Xiao20PRB} or a space symmetry related hidden order transition \cite{Zhao16np,Zhao17np}.
It has recently been proposed that the nonlinear Hall effect can also be used to probe the quantum critical point \cite{Facio18prl,Du18prl,Habib20prr} and N\'{e}el vector orientation in antiferromagnets \cite{Shao20prl}.
Various related phenomena have also been proposed, such as the gyrotropic Hall effect \cite{Konig19prb}, the Magnus Hall effect \cite{Papaj19prl}, and the nonlinear Nernst effect \cite{Su19prb,Zeng19prb}.

The nonlinear Hall effect has a quantum nature because of its connection with the Berry curvature dipole.
The Berry curvature can be regarded as a magnetic field in parameter space (e.g., momentum space).
It describes the bending of parameter spaces, arising from the geometric structure of quantum eigen states.
The Berry curvature dipole describes the dipole moment of the Berry curvature in momentum space \cite{Sodemann15prl}.
In addition, the nonlinear Hall effect is a quantum transport phenomenon near the $dc$ limit because of the extremely low frequency ($\sim$ 10 to 1000 Hz) of the input currents in experiments \cite{Ma19nat,Kang19nm,Dzsaber21PNAS,Qin2021CPL,Ho2021NE,Huang2020arXiv,Tiwari21NC,Kiswandhi21arXiv,He2021NC,Kumar2021NN}.
The importance of the quantum description of $dc$ quantum transports has been well acknowledged \cite{Lee85rmp}.
Despite its quantum nature, by far there are only semiclassical theories based on the Boltzmann equations under the relaxation time approximation \cite{Sodemann15prl,Du18prl,Du19nc,Xiao19prb,Nandy19prb,Isobe20sa,Konig19prb}.
There has been a tendency towards a quantum description of the nonlinear Hall effect \cite{Nandy19prb,Xiao19prb}.
A new side-jump contribution without semiclassical correspondence has also been discovered \cite{Xiao19prb}.
However, a systematic quantum theory of the nonlinear Hall effect that can explicitly describe the disorder effects is yet to be developed.

\begin{figure*}[htbp]
\centering
\includegraphics[width=1\textwidth]{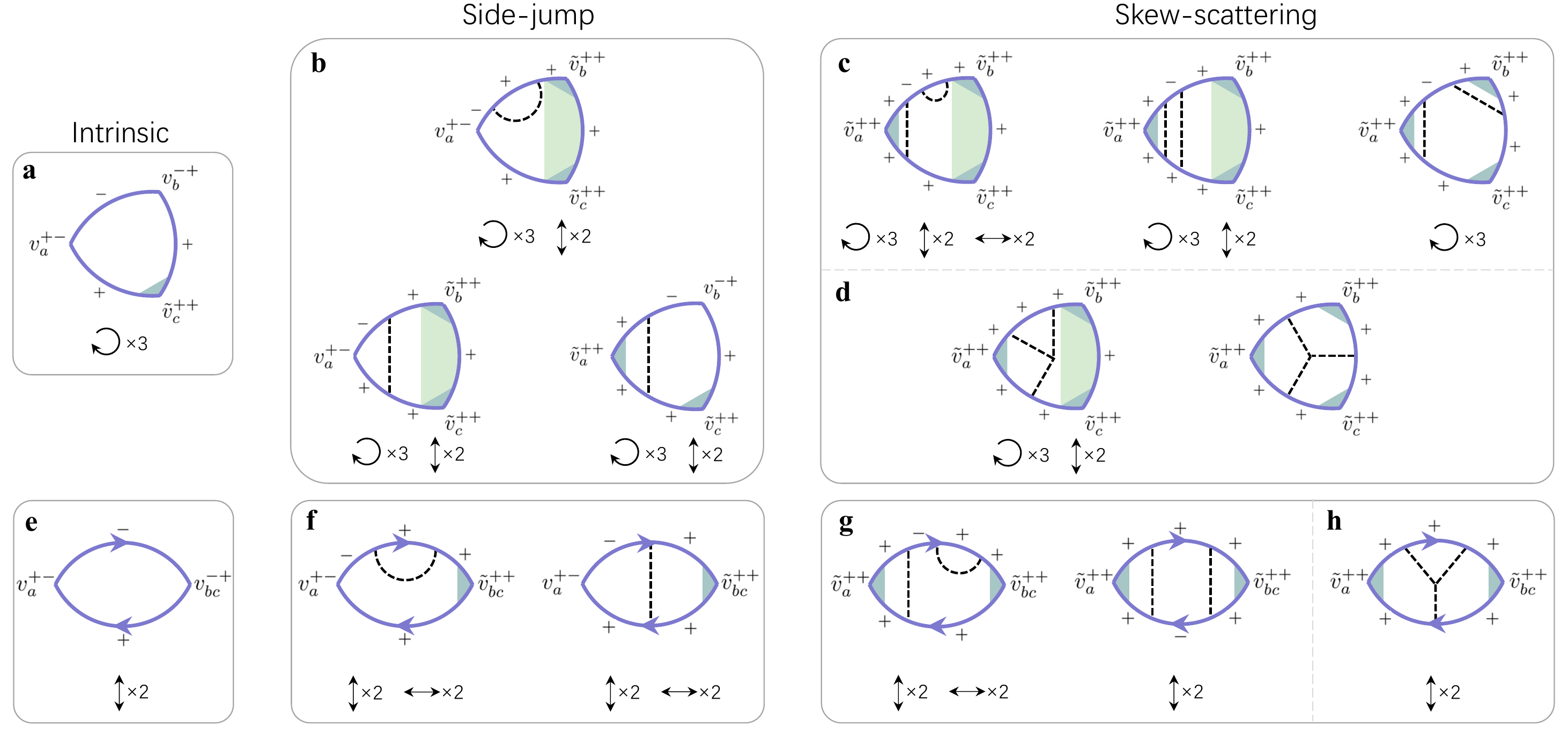}
\caption{\textbf{Feynman diagrams for the nonlinear Hall conductivity}.
The eigen bands of a generic two-band model are labeled as $\pm$ and the Fermi energy is assumed to cross the $+$ band.
\textbf{a}-\textbf{d} are the triangular diagrams and \textbf{e}-\textbf{h} are the two-photon diagrams.
These diagrams can be classified into intrinsic (\textbf{a} and \textbf{e}), side-jump (\textbf{b} and \textbf{f}), intrinsic skew-scattering (\textbf{c} and \textbf{g}), and extrinsic skew-scattering (\textbf{d} and \textbf{h}) contributions.
The solid lines stand for the Matsubara Green's function of the $+$ or $-$ band. The dashed lines represent the disorder scattering. The gray and green shadows represent the vertex and edge corrections, respectively.
The triangular diagrams can be labeled as both clockwise and anticlockwise arrows \cite{Kamenev95PRB,Flensberg95PRB,Narozhny16RMP,Parker19prb,Michishita21PRB},
which we count as one diagram.
The symbols $\circlearrowright\times3$, $\updownarrow\times2$ and $\leftrightarrow\times2$ represent the three-fold permutation, up-down and left-right reverse of the scattering kernel and accordingly the labels of the Green's functions and vertices.
These symbols give the number of each type of diagrams. For example, \textbf{a} has three diagrams by three-fold permutations of the \{$-$, $+$, $+$\} labels.
All of these diagrams can be found in Supplementary Figs.~{\color{blue}5}-{\color{blue}9} (triangular) and {\color{blue}11}-{\color{blue}15} (two-photon).}\label{Fig:Diagrams}
\end{figure*}

In this work, we construct a quantum theory for the nonlinear Hall effect using the diagrammatic technique.
Unlike the bubble diagrams of the linear-response theory, the quadratic responses are described by triangular and two-photon diagrams, representing two inputs and one output. We identify 69 Feynman diagrams that contribute to the leading nonlinear responses in the weak-disorder limit, including the intrinsic, side-jump, and skew-scattering contributions (Fig.~\ref{Fig:Diagrams}).
Quite different from nonlinear optics \cite{Sturman92book,Sipe00prb,Taguchi16prbrc,Chan17prbrc,Parker19prb}, 64 out of these diagrams account for the disorder effects, which are decisive for the electronic transport.
We formulate the diagrams for a generic two-band model and apply them to calculate the nonlinear Hall conductivity of a disordered 2D tilted Dirac model.
The general formulas obtained from the diagrammatic calculations can be directly adopted by the first-principles calculations.
According to the symmetry of the diagrams, we perform a symmetry analysis of the nonlinear Hall response tensor for all of the 32 point groups (see Table~\ref{Tab:Comparison} for 2D and Supplementary Table {\color{blue}2} for 3D).

\section{Results}
\textbf{Nonlinear response and Feynman diagrammatics}.
In response to $ac$ electric fields along the $b$ and $c$ directions, the nonlinear electric current along the $a$ direction can be formally written as (Supplementary Note 2)
\begin{eqnarray}\label{Eq:Define}
\mathrm{Re}[J^{(2)}_{a}(t)]&=&\xi_{abc}\mathcal{E}_{b}\mathcal{E}_{c}\cos[(\omega_{b}-\omega_{c})t]\nonumber\\
&&+\chi_{abc}\mathcal{E}_{b}\mathcal{E}_{c}\cos[(\omega_{b}+\omega_{c})t],
\end{eqnarray}
where $\{a,b,c\}\in\{x,y,z\}$, $\mathcal{E}_{b,c}$ and $\omega_{b,c}$ are the amplitudes and frequencies of the electric fields, respectively. For a mono-frequency electric field input, $\xi_{abc}$ and $\chi_{abc}$ are the zero- and double-frequency responses, respectively, and we should have $\xi_{abc}=\chi_{abc}$ by definition when $\omega_{b}=\omega_{c}=0$.
In experiments, it is more convenient to measure the double-frequency response, which is less sensitive to low-frequency noises, so we focus on $\chi_{abc}$.

\begin{figure*}[htbp]
\centering
\includegraphics[width=1\textwidth]{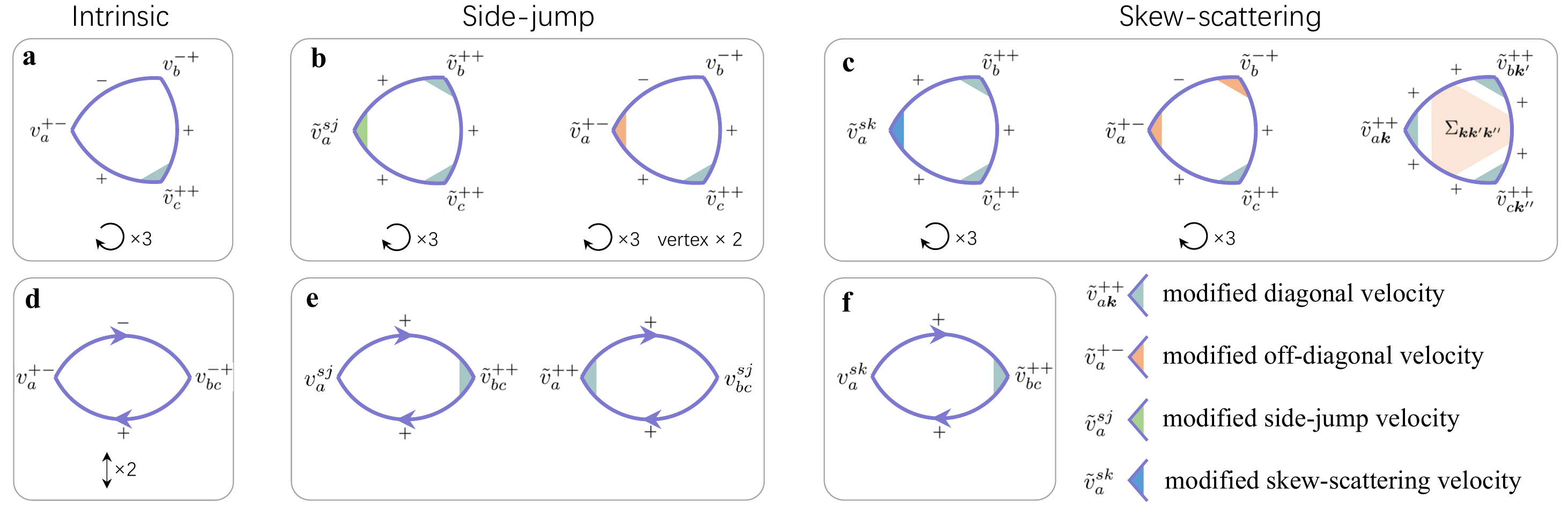}
\caption{\textbf{Simplified Feynman diagrams for the nonlinear Hall conductivity}.
The simplified triangular (\textbf{a}-\textbf{c}) and two-photon (\textbf{d}-\textbf{f}) diagrams after the redefinition of the vertex functions.
The gray shadow represents the modified diagonal velocity that has appeared in Fig.~\ref{Fig:Diagrams}, while the orange, green, and blue shadows represent the newly defined modified off-diagonal, side-jump, and skew-scattering velocities respectively. $\Sigma_{\bm{k}\bm{k}'\bm{k}''}$ is the indecomposable self-energy.
The symbols $\circlearrowright\times3$ and $\updownarrow\times2$ are the same as those in Fig.~\ref{Fig:Diagrams}, and vertex$\times2$ in \textbf{b} means that for the shown diagram one can label either $v^{+-}_{a}$ or $v^{-+}_{b}$ as the modified off-diagonal vertex.
Detailed definition of each quantity within the non-crossing approximation and all the simplified diagrams can be found in Supplementary Note 6.}\label{Fig:Reduce}
\end{figure*}

It has been pointed out that \cite{Parker19prb,Xu2019PRB}, in order to compute a gauge invariant non-linear conductivity up to quadratic order in electric fields, one should expand the vector potential $\bm{A}$ in the Peierls substituted Hamiltonian $\hat{\mathcal{H}}(\bm{k}+e\bm{A}/\hbar)$ up to the third order, where $-e$ is the electron charge.
These high-order terms are proportional to the tensor generalization ($\hat{v}_{ab}=\partial^{b}_{\bm{k}}\partial^{a}_{\bm{k}}\hat{\mathcal{H}}/\hbar^2$ and
$\hat{v}_{abc}=\partial^{c}_{\bm{k}}\partial^{b}_{\bm{k}}\partial^{a}_{\bm{k}}\hat{\mathcal{H}}/\hbar^3$) of the velocity operator $\hat{v}_{a}=\partial^{a}_{\bm{k}}\hat{\mathcal{H}}/\hbar$, where $\partial^{a}_{\bm{k}}\equiv\partial/\partial k_{a}$, $\hat{v}_{ab}$ and $\hat{v}_{abc}$ correspond to the two- and three-photon processes, respectively.
To achieve a divergent free description in the $dc$ limit, one needs to treat all the coupling vertices on the same footing.
After a lengthy calculation (Supplementary Note 2), one can obtain the quadratic conductivity in the $dc$ limit as \cite{Michishita21PRB} $\chi_{abc}=\chi^{I}_{abc}+\chi^{II}_{abc}+\chi^{III}_{abc}$, where
\begin{widetext}
\begin{eqnarray}
\chi^{I}_{abc}
&=&-\frac{e^3\hbar^2}{4\pi}\int[dk]\int^{\infty}_{-\infty}d\varepsilon
\frac{\partial f(\varepsilon)}{\partial\varepsilon}
\mathrm{Im}\Big\{\mathrm{Tr}\Big[\hat{v}_{a}\frac{\partial\hat{G}^{R}(\varepsilon)}{\partial\varepsilon}\hat{v}_{b}\hat{G}^{R}(\varepsilon)\hat{v}_{c}\hat{G}^{A}(\varepsilon)\Big]\Big\}
+b\leftrightarrow c,\label{Eq:chiI}
\\
\chi^{II}_{abc}
&=&-\frac{e^3\hbar^2}{8\pi}\int[dk]\int^{\infty}_{-\infty}d\varepsilon
\frac{\partial f(\varepsilon)}{\partial\varepsilon}
\mathrm{Im}\Big\{\mathrm{Tr}\Big[\hat{v}_{a}\frac{\partial\hat{G}^{R}(\varepsilon)}{\partial\varepsilon}\hat{v}_{bc}\hat{G}^{A}(\varepsilon)\Big]\Big\}
+b\leftrightarrow c,\label{Eq:chiII}
\\
\chi^{III}_{abc}
&=&-\frac{e^3\hbar^2}{8\pi}\int[dk]\int^{\infty}_{-\infty}d\varepsilon f(\varepsilon)\mathrm{Im}\Big\{
\mathrm{Tr}\Big\{\hat{v}_{a}\frac{\partial^{2}\hat{G}^{R}(\varepsilon)}{\partial\varepsilon^{2}}\hat{v}_{bc}\hat{G}^{R}(\varepsilon)
+2\hat{v}_{a}\frac{\partial}{\partial\varepsilon}\Big[\frac{\partial\hat{G}^{R}(\varepsilon)}{\partial\varepsilon}\hat{v}_{b}\hat{G}^{R}(\varepsilon)\Big]
\hat{v}_{c}\hat{G}^{R}(\varepsilon)\Big\}\Big\}
+b\leftrightarrow c,\label{Eq:chiIII}
\end{eqnarray}
\end{widetext}
$[dk]\equiv d^{n}\bm{k}/(2\pi)^{n}$ with $n$ for the dimensionality,
$\mathrm{Im}(\hat{\mathcal{O}})\equiv(\hat{\mathcal{O}}-\hat{\mathcal{O}}^{\dagger})/2i$ for an operator $\hat{\mathcal{O}}$,
$\hat{G}^{R/A}(\varepsilon)$ is the retarded/advanced Green's function,
and $f(\varepsilon)=1/\{1+\exp[(\varepsilon-\varepsilon_{\mathrm{F}})/k_B T]\}$ is the Fermi distribution with the Fermi energy $\varepsilon_{\mathrm{F}}$.
$\chi^{I}_{abc}$ and $\chi^{II}_{abc}$ are the Fermi surface contributions, where $\chi^{I}_{abc}$ describes the triangular diagrams while $\chi^{II}_{abc}$ describes the two-photon diagrams.
The diagrammatic representation of $\chi^{I}_{abc}$ and $\chi^{II}_{abc}$ are not standard but effective one obtained from analytical calculations (Supplementary Note 2).
$\chi^{III}_{abc}$ is the Fermi sea contribution, in which all the terms depend on the products of $\hat{G}^{R}$ only or $\hat{G}^{A}$ only.
It can be shown that terms in $\chi^{III}_{abc}$ are one order smaller than the leading terms in $\chi^{I}_{abc}$ and $\chi^{II}_{abc}$ in the weak-disorder limit \cite{Mahan1990}, we can hence neglect the Fermi sea contributions in the low-frequency transports.
By transforming into the eigenstate basis, we can describe different mechanisms of the nonlinear Hall conductivity explicitly within the diagrammatic approach.
In the weak-disorder limit, only the contributions in the leading order of the impurity concentration $n_{i}$ are important, and thus the diagrams are selected according to their $n_{i}$ dependence (see Methods).
The relevant diagrams for time-reversal symmetric systems are shown in Fig.~\ref{Fig:Diagrams}, which can be further classified into intrinsic, side-jump, and skew-scattering diagrams according to their correspondences in the semiclassical descriptions.

\begin{figure*}[htpb]
\centering
\includegraphics[height=0.52\textwidth]{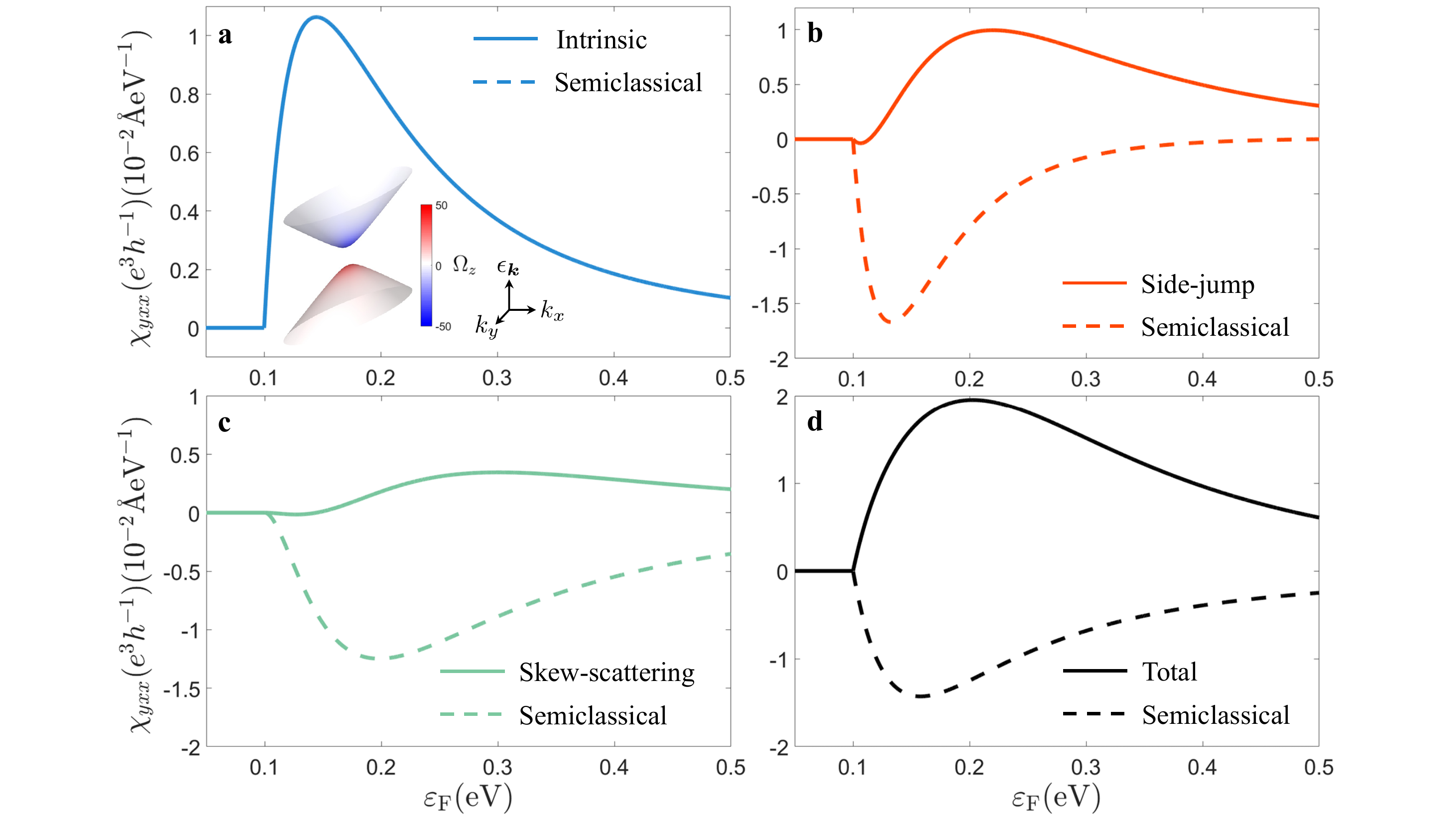}
\caption{\textbf{Nonlinear Hall conductivity of a 2D tilted Dirac model}.
(\textbf{a}-\textbf{d}) The intrinsic, side-jump, skew-scattering, and total contributions to the nonlinear Hall conductivity $\chi_{yxx}$ of the 2D tilted massive Dirac model [Eq.~(\ref{Eq:TiltedDirac})] as functions of the Fermi energy $\varepsilon_{F}$ at zero temperature. The semiclassical results (dashed lines) are also shown for comparison. The inset of \textbf{a} shows the two energy bands of the model. The color bar stands for the value of the Berry curvature.
The model parameters are $t=0.1~\mathrm{eV\cdot\AA}$, $v=1~\mathrm{eV\cdot\AA}$, $m=0.1~\mathrm{eV}$, $n_{i}V^{2}_{0}=10^2~\mathrm{eV}^2\cdot\mathrm{\AA}^2$ and $n_{i}V^{3}_{1}=10^4~\mathrm{eV}^3\cdot\mathrm{\AA}^4$.}\label{Fig:Conductivity}
\end{figure*}

\textbf{Generic model and disorder}.
We consider a generic two-band model as a building block for realistic systems
\begin{eqnarray}\label{Eq:model}
\hat{\mathcal{H}}=h_0+h_x\sigma_x+h_y\sigma_y+h_z\sigma_z,
\end{eqnarray}
where $\sigma_{x,y,z}$ are the Pauli matrices, $h_0$ and $h_{x,y,z}$ are functions of the wave vector $\bm{k}=(k_x,k_y,k_z)$.
The model describes two energy bands (denoted as $\pm$) with the band dispersions
$\varepsilon^{\pm}_{\bm{k}}=h_{0}\pm h_{\bm{k}}$, where $h_{\bm{k}}\equiv(h^{2}_{x }+h^{2}_{y }+h^{2}_{z })^{1/2}$.
The disorder is modeled as $\delta$-function scatters
$V_{imp}(\bm{r})=\sum_{i}V_{i}\delta(\bm{r}-\bm{R}_{i})$ with a random distribution $\bm{R}_{i}$ and the disorder strength $V_{i}$ satisfying $\langle V_{i}\rangle=0$, $\langle V^{2}_{i}\rangle=V^{2}_{0}$, and $\langle V^{3}_{i}\rangle=V^{3}_{1}$, where $\langle...\rangle$ means the ensemble average over disorder configurations.
Up to the leading order, the disorder scattering has two types of correlation: one correlates two scattering events (Gaussian disorder distribution),
and the other correlates three scattering events (non-Gaussian disorder distribution).

The above considerations allow us to identify the physical mechanism of each diagram (see Methods) and derive the expressions of the nonlinear Hall conductivity for the generic two-band model (Supplementary Note 3 and 4).
The intrinsic part up to the leading order is only contributed by Fig.~\ref{Fig:Diagrams} \textbf{a} and still proportional to the Berry curvature dipole, its expression within the quantum theory is
\begin{eqnarray}
\chi^{in}_{abc}
&=&-\frac{e^3}{4\hbar}\int[dk]\tau^{+}_{\bm{k}}f'(\varepsilon^{+}_{\bm{k}})
\varepsilon^{abd}\Omega^{+}_{d\bm{k}}\tilde{v}^{++}_{c\bm{k}}+b\leftrightarrow c,
\end{eqnarray}
where $f'(\varepsilon)\equiv\partial f(\varepsilon)/\partial\varepsilon$, $\varepsilon^{acd}$ is the Levi-Civita anti-symmetric tensor, $\tau^{+}_{\bm{k}}$ is the scattering time, $\Omega^{+}_{a\bm{k}}$ is the Berry curvature, and $\tilde{v}^{++}_{a\bm{k}}$ is the vertex-corrected diagonal velocity.
As we do not assume a detailed vertex-correction form of $\tilde{v}^{++}_{a\bm{k}}$ that corresponds to the gray shadow in Fig.~\ref{Fig:Diagrams} \textbf{a}, many possible quantum corrections can in principle be accounted in forms of $\tilde{v}^{++}_{a\bm{k}}$.
More strikingly, the side-jump and skew-scattering parts include qualitatively new contributions that go beyond the semiclassical description (Supplementary Note 3 and 4).
Especially, the contribution from the two-photon diagrams is related to the leading $\omega$-dependent part of the vertex correction, which is a peculiar feature of the nonlinear transports.
As the semiclassical description of the nonlinear Hall effect is obtained by generalizing the modern semiclassical theory of the anomalous Hall effect \cite{Sinitsyn07jpcm} to the nonlinear response regime, this qualitative inconsistency indicates that a proper semiclassical theory of the nonlinear Hall effect warrants some special considerations.

Nevertheless, we find that the quantum theory does not modify the scaling law because the scaling behavior is determined by the order of the disorder-dependence of each mechanism \cite{Du19nc,Tiwari21NC}, while so far we find that the disorder dependence is the same for the semiclassical and quantum theories.

\begin{table*}[htbp]
\centering
\caption{\textbf{Nonzero nonlinear Hall response elements in 32 point groups (2D).} The matrices are defined in Eq.~(\ref{Eq:2D-relation}). The elements that exist in $\chi^{ex}$ but vanish in $\chi^{in}$ are highlighted.
The $C_{n}$ axis is assumed in order as the $z$-, $x$- and $y$-axis, the mirror plane $\sigma_{v}$ is assumed in order as the $yz$- and $zx$-plane and the mirror plane $\sigma_{h}$ is assumed as the $xy$-plane.}\label{Tab:Comparison}
\begin{ruledtabular}
\begin{tabular}{cccc}
Class/Group & Extrinsic contribution & Intrinsic contribution \\
\hline
Triclinic/$C_{1}$ & $\left(\begin{array}{ccccccc}
                       {\color{red}\chi^{ex}_{xxx}} & \chi^{ex}_{xxy} & \chi^{ex}_{xyy} \\
                       \chi^{ex}_{yxx} & \chi^{ex}_{yyx} & {\color{red}\chi^{ex}_{yyy}} \\
                     \end{array}
                   \right)$ &
                   $\left(
                     \begin{array}{ccccccc}
                       0 & -\chi^{in}_{yxx} & \chi^{in}_{xyy} \\
                       \chi^{in}_{yxx} & -\chi^{in}_{xyy} & 0 \\
                     \end{array}
                   \right)$
                   \\
\hline
Monoclinic/$C_{1v}$ & $\left(
                     \begin{array}{ccccccc}
                       0 & \chi^{ex}_{xxy} & 0  \\
                       \chi^{ex}_{yxx} & 0 & {\color{red}\chi^{ex}_{yyy}} \\
                     \end{array}
                   \right)$ &
                   $\left(
                     \begin{array}{ccccccc}
                       0 & -\chi^{in}_{yxx} & 0 \\
                       \chi^{in}_{yxx} & 0 & 0\\
                     \end{array}
                   \right)$
                   \\
\hline
$\begin{array}{ccccccc} \text{Trigonal}/C_{3} \\ \text{Hexagonal}/C_{3h}\end{array}$ & $\left(
                     \begin{array}{ccccccc}
                       {\color{red}\chi^{ex}_{xxx}} & {\color{red}-\chi^{ex}_{yyy}} & {\color{red}-\chi^{ex}_{xxx}}\\
                       {\color{red}-\chi^{ex}_{yyy}} & {\color{red}-\chi^{ex}_{xxx}} & {\color{red}\chi^{ex}_{yyy}} \\
                     \end{array}
                   \right)$ &
                   $0$
                   \\
\hline
$\begin{array}{ccccccc} \text{Trigonal}/C_{3v} \\ \text{Hexagonal}/D_{3h}\end{array}$ & $\left(
                     \begin{array}{ccccccc}
                       0 & {\color{red}-\chi^{ex}_{yyy}} & 0 \\
                       {\color{red}-\chi^{ex}_{yyy}} & 0 & {\color{red}\chi^{ex}_{yyy}} \\
                     \end{array}
                   \right)$ &
                   $0$
                   \\
\hline
Trigonal/$D_{3}$ & $\left(
                     \begin{array}{ccccccc}
                       {\color{red}\chi^{ex}_{xxx}} & 0 & {\color{red}-\chi^{ex}_{xxx}} \\
                       0 & {\color{red}-\chi^{ex}_{xxx}} & 0  \\
                     \end{array}
                   \right)$ &
                   $0$
\end{tabular}
\end{ruledtabular}
\end{table*}

\textbf{Simplified representation of the Feynman diagrams}.
For a better understanding of the side-jump and skew-scattering diagrams, it is desirable to introduce some quantities that can simplify the diagrammatic representation.
The 16 disorder-related diagrams of the anomalous Hall effect can be simplified into 3 by introducing the side-jump and skew-scattering velocities $v^{sj}_{a}$ and $v^{sk}_{a}$.
As the two-photon diagrams of the nonlinear Hall effect share the same Green's function parts as those of the anomalous Hall effect, these diagrams can also be simplified by introducing $v^{sj}_{a}$ and $v^{sk}_{a}$ but with an additional tensor generalization of the side-jump velocity $v^{sj}_{ab}$ as shown in Fig.~\ref{Fig:Reduce} \textbf{e} and \textbf{f}.
To simplify the triangular diagrams, we have introduced the modified off-diagonal, side-jump, and skew-scattering velocities $\tilde{v}^{+-}_{a}$, $\tilde{v}^{sj}_{a}$, and $\tilde{v}^{sk}_{a}$, respectively.
By introducing these quantities we can simplify the 46 disorder-induced triangular diagrams into 16 as shown in Fig.~\ref{Fig:Reduce} \textbf{b} and \textbf{c} (Supplementary Note 6).

Other than reducing the number of the diagrams, the simplified representation of the diagrams highlights the qualitative difference between the nonlinear Hall diagrams and the anomalous Hall ones.
Although $\tilde{v}^{sj}_{a}$ and $\tilde{v}^{sk}_{a}$ can be considered as the generalization of $v^{sj}_{a}$ and $v^{sk}_{a}$, the disorder modification complicates the quantum results so much that they are very different from their semiclassical counterparts.
In addition, the $\tilde{v}^{+-}_{a}$ related diagrams obviously do not have any linear counterpart.
More interestingly, these simplified diagrams show similar structures as the intrinsic triangular diagrams, although the $\tilde{v}^{+-}_{a}$ related skew-scattering diagrams do not contribute to the nonlinear Hall conductivity within our simple considerations.
Another important difference comes from the diagrams with the indecomposable self-energy $\Sigma_{\bm{k}\bm{k}'\bm{k}''}$.
Although it also vanishes in our consideration, this type of diagrams can be important once we go beyond the non-crossing approximation.
This simplified representation of the nonlinear Hall diagrams is general and is not restricted by models or approximations.

\textbf{Application to the 2D tilted Dirac model}.
For an intuitive estimate of the quantum contributions, we apply the diagrams to calculate the nonlinear response for the 2D tilted Dirac model, whose Hamiltonian can be obtained by letting
\begin{eqnarray}\label{Eq:TiltedDirac}
h_{0}=tk_{x},\ h_{x}=vk_{x},\ h_{y}=vk_{y},\ h_{z}=m
\end{eqnarray}
in Eq.~(\ref{Eq:model}) with the model parameters $t$, $v$ and $m$. This is the minimal model of the nonlinear Hall effect because it has strong Berry curvature and has no inversion symmetry \cite{Sodemann15prl,Ma19nat,Du18prl}. $t/v$ measures the tilt of the Dirac cone along the $x$ direction, which breaks inversion symmetry. $2m$ is the band gap \cite{Du18prl}.
A single 2D Dirac cone does not have time-reversal symmetry. Time-reversal symmetry is satisfied by including its time-reversal partner ($m\rightarrow -m$, $t\rightarrow -t$) at opposite regions of the Brillouin zone \cite{Ma19nat}, which contributes the same nonlinear Hall response by symmetry.

With the help of the effective diagrammatics, the nonlinear Hall conductivity $\chi_{yxx}$ of the 2D Dirac model at zero temperature can be obtained (Supplementary Note 8), as shown in Fig.~\ref{Fig:Conductivity}.
Within the non-crossing approximation, the intrinsic contribution from the quantum theory is identical to the result from the semiclassical theory \cite{Sodemann15prl} (Fig.~\ref{Fig:Conductivity} \textbf{a}).
However, the side-jump and skew-scattering contributions calculated by the quantum theory demonstrate opposite signs, compared to the semiclassical results (Fig.~\ref{Fig:Conductivity} \textbf{b} and \textbf{c}).
As a result, the nonlinear Hall conductivity also has opposite signs for the semiclassical and quantum theories (Fig.~\ref{Fig:Conductivity} \textbf{d}).
Different from the semiclassical results, the total quantum result shares the same sign and similar line shape with the intrinsic contribution but with a greater magnitude.
The sign change from the semiclassical theory to the quantum theory is partially supported by another work beyond the semiclassical theory, where a tendency of the sign change is observed as the quantum description comes in  \cite{Nandy19prb}.
Our calculation results then provide an explanation on the fact that although the scaling experimental results indicate a comparable intrinsic and disorder-induced contributions \cite{Kang19nm}, the qualitative feature of the nonlinear Hall effect can still be well described by the Berry curvature dipole in bilayer WTe$_{2}$ \cite{Ma19nat,Du18prl}.

Our quantitative results come from a case study, which unnecessarily includes certain approximations on the methods and models.
As the competition between the intrinsic and extrinsic mechanisms may be different from case to case, future studies are needed to reveal the possible general rules.
Nevertheless, our calculation clearly shows that the quantum description is very important for the nonlinear Hall effect, especially when the disorder effects are relevant.

\textbf{Symmetry aspects of the nonlinear response}.
According to our diagrammatic results, the disorder-induced extrinsic contribution to the quadratic nonlinear conductivity is a rank-three tensor with the constraint $\chi^{ex}_{abc}=\chi^{ex}_{acb}$, while the intrinsic contribution related to the Berry curvature dipole has extra antisymmetric properties under the label exchanges $a\leftrightarrow b$ or $a\leftrightarrow c$, as described by
\begin{eqnarray}\label{Eq:From}
\chi^{ex}_{abc}&=&T_{abc}+T_{acb},
\\
\chi^{in}_{abc}&=&\varepsilon_{abd}T_{cd}+\varepsilon_{acd}T_{bd},
\end{eqnarray}
where $T_{ab}$ and $T_{abc}$ are rank-two and rank-three tensors, respectively.
The extrinsic response tensor $\chi^{ex}$ has more nonzero elements than $\chi^{in}$.
More importantly, the different symmetry properties under the exchange of labels impose different constraints on the elements of the nonlinear response tensor.

For a complete investigation, we check both the nonzero elements of $\chi^{ex}$ and $\chi^{in}$ for all of the 32 point groups (Supplementary Note 9). The results for 2D systems are summarized in Table \ref{Tab:Comparison}, where the matrix is defined as
\begin{eqnarray}\label{Eq:2D-relation}
\left(\begin{array}{c}
  J_{x}\\
  J_{y}
\end{array}\right)&=&\left(
                     \begin{array}{cccc}
                       \chi_{xxx} & \chi_{xxy} & \chi_{xyy} \\
                       \chi_{yxx} & \chi_{yyx} & \chi_{yyy} \\
                     \end{array}
                   \right)\left(\begin{array}{c}
                   E^{2}_{x} \\
                   2E_{x}E_{y} \\
                   E^{2}_{y} \\
                   \end{array}\right),
\end{eqnarray}
and we have highlighted the elements that exist in $\chi^{ex}$ but vanish in $\chi^{in}$.
The results for 3D systems can be found in Supplementary Table {\color{blue}2}.
These elements are contributed by the disorder effects, and thus represent the Berry curvature irrelevant nonlinear Hall response.
In some point groups, such as $C_{3}$, $C_{3h}$, $C_{3v}$, $D_{3h}$, and $D_{3}$ in 2D, the Berry-curvature-dipole-related $\chi^{in}$ vanishes, but $\chi^{ex}$ survives.
For 3D systems, the point groups that support the pure disorder-induced nonlinear Hall effect are $T$, $T_{d}$, $C_{3h}$, and $D_{3h}$.
Therefore, the nonlinear Hall effect observed in systems with these point groups can only be induced by disorder.

\section{Methods}
\textbf{Diagram construction}.
In the weak-disorder limit, the diagrams of leading contribution are constructed according to their dependence on the impurity concentration $n_{i}$ \cite{Mahan1990}.
For systems with time-reversal symmetry, the leading contribution to the nonlinear transport is of order $n^{-1}_{i}$, which can be obtained by adding non-ladder-type scattering events to the simplest triangular and two-photon diagrams \cite{Sinitsyn07prb}.
The resulting diagrams of order $n^{-1}_{i}$ within the non-crossed approximation are shown in Fig.~\ref{Fig:Diagrams} \textbf{a}-\textbf{h}, which include intrinsic, side-jump, intrinsic and extrinsic skew-scattering contributions.
A complete summary of all the 69 diagrams for each contribution can be found in Supplementary Figs.~{\color{blue}5}-{\color{blue}9} (triangular) and {\color{blue}11}-{\color{blue}15} (two-photon).

The classification of these diagrams is carried out via finding the characteristic physical quantities.
For the intrinsic contribution (Figs.~\ref{Fig:Diagrams} \textbf{a} and \textbf{e}), the characteristic quantity is the Berry curvature.
For the side-jump contribution (Figs.~\ref{Fig:Diagrams} \textbf{b} and \textbf{f}), the characteristic quantity is $v_{a\bm{k}}^{+-}\langle V^{-+}_{\bm{k}\bm{k}'}V^{++}_{\bm{k}'\bm{k}}\rangle$ with $V^{-+}_{\bm{k}\bm{k}'}\equiv\langle u^{-}_{\bm{k}}|{V}_{imp}|u^{+}_{\bm{k}'}\rangle$, which represents an off-diagonal scattering process.
The skew-scattering contribution contains two categories as intrinsic (Figs.~\ref{Fig:Diagrams} \textbf{c} and \textbf{g}) and extrinsic (Figs.~\ref{Fig:Diagrams} \textbf{d} and \textbf{h}) skew-scattering according to their characteristic scattering processes (Supplementary Note 7).
The first one is from the leading asymmetric scattering contribution due to the Gaussian disorder within the non-crossing approximation, which is characterised by $\langle V^{-+}_{\bm{k}\bm{k}'}V^{++}_{\bm{k}'\bm{k}}\rangle\langle V^{+-}_{\bm{k}''\bm{k}}V^{++}_{\bm{k}\bm{k}''}\rangle$.
The second one is from the leading asymmetric scattering contribution due to the non-Gaussian disorder, which is characterised by
$\langle V^{++}_{\bm{k}\bm{k}'}V^{++}_{\bm{k}'\bm{k}''}V^{++}_{\bm{k}''\bm{k}}\rangle$.

\textbf{Remarks on the effective diagrammatics}.
The effective diagrammatic representations of $\chi^{I}_{abc}$ and $\chi^{II}_{abc}$ provide us an approach for a quantum description of the nonlinear Hall effect, although its construction is not so straight forward and indicates the importance of the multi-photon processes even for systems with a linear $\bm{k}$-dependent Hamiltonian.
Because the establishment of Eq.~(\ref{Eq:chiI})-(\ref{Eq:chiIII}) requires that all the coupling vertices ($\hat{v}_{a}$, $\hat{v}_{ab}$, $\hat{v}_{abc}$) should be finite,
a correct interpretation of the effective diagrammatic theory is that we should first include the multi-photon coupling vertices to obtain the general expression in the $dc$ limit, and then turn to the low-energy effective Hamiltonian for detailed calculations.
Alternatively, a direct calculation of the triangular diagrams would lead to unphysical results in the $dc$ limit (Supplementary Note 5).
A possible reason for this puzzle is that, the linear $\bm{k}$-dependent Hamiltonian is only a low-energy effective description but there are always multi-photon coupling vertices for the Bloch bands.
An accurate description of the nonlinear Hall effect may require the full-band information even though it is only a Fermi surface effect.

\textbf{Code availability}
The code that is deemed central to the conclusions are available from the corresponding author upon reasonable request.

\textbf{Data availability}
The data that support the plots within this paper and other findings of this study are available from the corresponding author upon reasonable request.

\section{Acknowledgments}
We thank helpful discussions with Huimei Liu, Suyang Xu, Alex Levchenko and Antti-Pekka Jauho.
This work was supported by the National Natural Science Foundation of China (12004157, 11974249, and 11925402), the National Basic Research Program of China (2015CB921102), the Strategic Priority Research Program of Chinese Academy of Sciences (XDB28000000), the Natural Science Foundation of Shanghai (Grant No. 19ZR1437300), Guangdong province (2020KCXTD001 and 2016ZT06D348), Shenzhen High-level Special Fund (G02206304 and G02206404), and the Science, Technology and Innovation Commission of Shenzhen Municipality (ZDSYS20170303165926217, JCYJ20170412152620376, and KYTDPT20181011104202253).
The numerical calculations were supported by Center for Computational Science and Engineering of Southern University of Science and Technology.

\section{Author contributions}
Z.Z.D. performed the calculations with assistance from C.M.W., H.-P.S., and H.-Z.L..
Z.Z.D. and H.-Z.L. wrote the manuscript with assistance from C.M.W., H.-P.S., and X.C.X..
H.-Z.L. and X.C.X. supervised the project.

\section{Competing financial interests}
The authors declare no competing interests.


\end{document}